\documentclass{article}

\usepackage{amssymb,amsfonts,amsmath,stmaryrd}
\usepackage{cite,enumerate,float,indentfirst}
\usepackage{color}

\def\be{\begin{eqnarray}}
\def\ee{\end{eqnarray}}
\def\nn{\nonumber}

\def\p{\partial}

\def\S{{\cal S}}
\def\CHI{{\cal S}}

\definecolor{red}{rgb}{1,0,0}
\definecolor{orange}{rgb}{1,0.5,0}
\definecolor{violet}{rgb}{0.7,0,1}

\hoffset= - 1.0in         

\begin{document}

\hfill ITEP/TH-23/18

\hfill IITP/TH-13/18

\bigskip

\centerline{\Large{An analogue of Schur functions for the plane partitions
}}

\bigskip

\centerline{A.Morozov}

\bigskip

\centerline{\it ITEP \& IITP, Moscow, Russia}

\bigskip

\centerline{ABSTRACT}

\bigskip

{\footnotesize
An attempt is described to extend the notion of Schur functions from Young diagrams
to plane partitions.
The suggestion is to use the recursion in the partition size,
which is easily generalized and deformed.
This opens a possibility to obtain Macdonald polynomials
by a change of recursion coefficients and taking appropriate
limit from three to two dimensions --
though details still remain to be worked out.
Another perspective is opened by the observation of a rich non-abelian structure,
extending that of commuting cut-and-join operators,
for which the discovered 3-Schurs are the common eigenfunctions.
}

\bigskip

\bigskip

\section{Introduction}

Characters are the central objects in physical applications of group theory,
because they describe invariant objects, which take values in {\it numbers},
still are often sufficient to capture important properties of correlators,
amplitudes and partition functions.
Especially interesting from this point of view is reformulation \cite{MAMOchars1}
of matrix models in terms of remarkable identity
\be
\Big<{\rm character}\Big> \ = \ character
\label{charchar}
\ee
where "characters" at the two sides are basically the same Schur functions,
only of different arguments -- quantum fields at the l.h.s.
and couplings (including matrix sizes) at the r.h.s.
This result reflects {\it super}integrability (a combination of ordinary
KP integrability and Ward-Virasoro constraints, reviewed in \cite{UFN3})
and is closely related to combinatorial treatment of matrix models
in \cite{MAMOchars2}, see \cite{MMcharreview} for details and references.
In \cite{IMMchar}, following the general logic of non-linear algebra
\cite{NLA}, this relation was extended from matrix to tensor models,
where one can find a very nice tensorial analogue of Schur functions,
despite most group theory structures are lost.
Another extension \cite{MPSh} is to {\it discrete} matrix models,
where ordinary integrals become Jackson sums and Schur functions
are substituted by Macdonald polynomials.
However, in this direction one naturally wants to go further --
to full-fledged generalization from Young diagrams to plane partitions
and from matrix to {\it network} models \cite{network}, AGT-related
to $6d$ SYM theories.
An important step of this kind was made in \cite{Z3pols},
but fully-$3d$ formulation was not quite achieved.
In this paper we make a kind of a complementary attempt --
from another side.
Hopefully, the two approaches can be unified and provide a much
better understanding.
In this paper we concentrate on ideally-symmetric $3d$ extension
of Schur functions
and only comment on the way to  Macdonald deformation.

\bigskip

Schur functions $S_\lambda\{p_k\}$
are labeled by Young diagrams (integer partitions) $\lambda$
and {\it therefore} depend on the one-parametric family of time-variables $p_k$,
which are the variables in the corresponding partition function.
Indeed, the states are $\Big|\{m\}\Big> = \prod_{k=1}^\infty p_k^{m_k}$,
and their generating function is
\be
\sum_{\{m\}} \prod_{k=1}^\infty p_k^{m_k}q^{km_k} = \prod_{k=1}^\infty \frac{1}{1-p_kq^k},
\ee
so that  the number of states is described by
\be
\sum_{\{m\}} \#_{\{m\}}\cdot q^{\sum_k km_k} =
\prod_{k=1}^\infty \frac{1}{1-q^k}
\ee
Their 3d-extension analogues ${\cal S}_\pi\{p_k^{(i)}\}$ should be
labeled by $3d$ diagrams (plane partitions) $\pi$ and thus depend on the
{\it triangular} set of time-variables $p_k^{(i)}$ with $i=1,\ldots,k$,
see \cite{Z3pols}.
Indeed, now the states are
$\Big|\{m\}\Big> = \prod_{1\leq i\leq k}^\infty (p_k^{(i)})^{m_k^{(i)}}$,
with the generating function
\be
\sum_{\{m\}} \prod_{1\leq i\leq k}^\infty (p_k^{(i)})^{m_k^{(i)}}(q^kT^i)^{m_k^{(i)}} =
\prod_{1\leq i\leq k}^\infty \frac{1}{1-p_k^{(i)}q^kT^i}
\ee
Then for the number of states we get
\be
\sum_{\{m\}} \#_{\{m\}}\cdot \prod_{1\leq i\leq k} (q^kT^k)^{m_k^{(i)}} =
\prod_{1\leq i\leq k}^\infty \frac{1}{1-q^kT^i}
\ee
and for $T=1$ this becomes the well known MacMahon function
\be
\prod_{k=1}^\infty \frac{1}{(1-q^k)^k} = 1+q+3q^2+6q^3+13q^4+24q^5+48q^6+86q^7 + \ldots
\label{MMah}
\ee

Explicit definition of Schur functions is usually a two-step process --
first one defines Schur polynomials for symmetric representations $[n]$
and then construct generic Schur functions via determinant formulas:
\be
e^{\sum_n \frac{1}{n}p_nz^n} = \prod_a (1-x_az) = \sum_n S_{[n]}\{p\}z^n
\ee
\be
S_\lambda = \det_{i,j\leq l_\lambda } S_{[\lambda_i-i+j]}
= \frac{\det_{a,b\leq l_\lambda+1} x_a^{\lambda_b-b+2}}{\Delta(x)}
\ee
Miwa transform relates $p$ and $x$ variables:
\be
p_n = \sum_a x_a^n
\ee
The true meaning of this procedure  is somewhat obscure,
it looks intimately related to the structure of fundamental representations,
associated with antisymmetrization (determinants) along the rows of the
Young diagrams -- and thus is not easy to generalize.

Alternative approach makes use of the generalized cut-and-join operators
from \cite{MMN1}, which are differential operators in $p$ or $x$:
\be
\hat W_R S_\lambda\{p\} = \psi_R(\lambda) S_\lambda\{p\}
\label{WopsMMN}
\ee
The point is that Schur functions are their common eigenfunctions,
and eigenvalues $\psi_R(\lambda)$  are  characters of symmetric group.
This relation is one of explicit realizations of the Schur-Weyl duality.
It is straightforward to deform from Schur to Macdonald polynomials,
however, generalization to plane partitions is not so easy,
because obscure is the appropriate substitute of the symmetric group.

Thus to proceed we need still another description of Schur functions --
which would be generalizable.
Such option is provided by the "path integral" (evolution) recursion,\footnote{
See \cite{Ded} for a renewed interest to this old viewpoint \cite{Segal,At,MR}.
In this context eq.(\ref{recSchur}) can be considered as an attempt (not yet fully
successful)
to build a quantum (topological) field theory, underlying the theory of symmetric
polynomials.
From another viewpoint,
(\ref{recSchur}) provides a  viable  deformation
of {\it locality} to the discrete space of partitions.
}
which involves skew characters
\be
\boxed{
S_{\lambda/\mu}\{p+p'\} = \sum_{\rho:\ \mu\subset\rho\subset\lambda}
S_{\lambda/\rho}\{p\}S_{\rho/\mu}\{p'\}
}
\label{recSchur}
\ee
Despite a seeming asymmetry, the r.h.s. is in fact symmetric under the permutation
of the sets $\{p\}$ and $\{p'\}$.
Here we use as input the definition of skew characters --
as linear decompositions of characters themselves,
i.e.  in this approach we {\it postulate} that
\be
S_{[n]/[m]} = S_{[n-m]},\ \ \ \ \ \ \ \
S_{[n,1]/[m]} = S_{[n-m+1]} + S_{[n-m,1]},\ \ \ \ \ \ \ \
\ldots
\ee
The point is that these relations can be interpreted as some basic
property of Young diagrams -- and then the problem is to derive what
are the associated characters $S_\mu$.
Then one can look for appropriate generalization of cut-and-join operators.
Our task in this paper is to show how all this works.

\section{Schur functions from recursion \label{recur}}

Solutions to the equation (\ref{recSchur}), even graded,
do not specify dependence on the "highest" $p$-variables
at each step.
For example:
\vspace{-0.2cm}
\be
\Delta S_{[1]} \equiv
S_{[1]}\{p+p'\} - S_{[1]}\{p\}-S_{[1]}\{p'\}=0 \ \Longrightarrow\ \ S_{[1]}=\alpha_{[1]} p_1
\ee
with arbitrary $\alpha_{[1]}$, which can be absorbed into rescaling of $p_1$,
thus we put $\alpha_{[1]}=1$.
Next,
\be
\Delta S_{[2]} \equiv  S_{[2]}\{p+p'\}-S_{[2]}\{p\}-S_{[2]}\{p'\} =
S_{[1]}\{p\}S_{[1]}\{p'\}
\ &\Longrightarrow \ \ \ S_{[2]}\{p\} = \frac{\alpha_{[2]}p_2 +
p_1^2}{2}
\nn \\
\Delta S_{[1,1]} \equiv  S_{[1,1]}\{p+p'\}-S_{[1,1]}\{p\}-S_{[1,1]}\{p'\}
= S_{[1]}\{p\}S_{[1]}\{p'\}
\ &\Longrightarrow \ \ S_{[1,1]}\{p\} = \frac{\alpha_{[1,1]}p_2 +p_1^2}{2}
\label{Schur2}
\ee
Further, in obvious notation:
\be
\Delta S_{[3]}\{p,p'\} = S_{[2]}\otimes S_{[1]}+S_{[1]}\otimes S_{[2]}
\ \ \Longrightarrow \ \ S_{[3]}\{p\} = \frac{2\alpha_{[3]}p_3 +
3\alpha_{[2]} p_2p_1
+  p_1^3}{6}
\nn \\
\Delta S_{[2,1]}\{p,p'\}
= (S_{[2]}+S_{[1,1]}) \otimes S_{[1]}+S_{[1]}\otimes(S_{[2]}+S_{[1,1]})
=\frac{\alpha_{[2]}+\alpha_{[1,1]}}{2}(p_{2}p_1' + p_1p_2')
+ \frac{1}{2}(p_1^2p_1'+p_1p_1'{\!^2})\!\!\!\!\!\!
\nn \\
\Longrightarrow \ \ S_{[2,1]}\{p\} = \frac{2\alpha_{[2,1]}p_3 +
3(\alpha_{[2]}+\alpha_{[1,1]})p_2p_1
+  p_1^3}{6}
\nn \\
\Delta S_{[1,1,1]}\{p,p'\} = S_{[1,1]}\otimes S_{[1]}+S_{[1]}\otimes S_{[1,1]}
\ \ \Longrightarrow \ \ S_{[1,1,1]}\{p\} = \frac{2\alpha_{[1,1,1]}p_3 +
3\alpha_{[1,1]} p_2p_1
+  p_1^3}{6}
\ee
and so on.
We introduced $\alpha$-parameters so that they are plus-minus unities or zeroes
for the true   Schur functions --
but, as we see, they are not fully restricted by (\ref{recSchur}).

There can be different ways to impose the further restrictions,
which specify $\alpha$'s,
the simplest one is to request orthogonality:
\vspace{-0.2cm}
\be
\hat S_R \cdot S_{R'} \sim \delta_{R,R'}
\ee
where $|R|=|R'|$ and
\vspace{-0.3cm}
\be
\hat S_R := S_{R}\left\{k\frac{\p}{\p p_k}\right\}
\label{duaSchur}
\ee
The choice of duality transform $p_k\longrightarrow k\frac{\p}{\p p_k}$
is dictated by  the properties of Schur functions, but
we will keep it intact for plain partitions --
though there is no clear motivation for  this.
At the same time, this transform is sensitive to $q,t$-transformation
from Schur to Macdonald functions.

One of the basic properties of Schur functions is that they are the
common eigenfunctions of an infinite commuting $W_Q$ operators (\ref{WopsMMN}),
of which the simplest  is the celebrated cut-and-join
\be
\hat W_{[2]} = \frac{1}{2}\sum_{a,b=1}^\infty \left((a+b)p_ap_b\frac{\p}{\p p_{a+b}}
+ abp_{a+b}\frac{\p}{\p p_a\p p_b}\right) =
p_1^2\frac{\p}{\p p_2}+\frac{p_2}{2}\frac{\p^2}{\p p_1^2} + \ldots
\label{W2Schur}
\ee
where dots stand for the terms containing $p_n$ or $\p_n$ with $n\geq 3$,
which do not contribute at the level $|R|=2$.

\section{Macdonald recursion
\label{MacD}}

Macdonald polynomials \cite{MD} also satisfy (\ref{recSchur}), but coefficients
are no longer unities -- they get $q,t$-deformed:
\be
\Delta M_{[n]} = \sum_{m=1}^{n-1}  \frac{<n>!}{<m>!<n-m>!} \cdot  M_{[m]}\otimes M_{[n-m]}
\ee
where $<n>\, = \frac{[n](1-t)}{1-q^{n-1}t}$ and $[n]=\frac{1-q^{n }}{1-q}$.
For antisymmetric representations Macdonald polynomials coincide with Schur functions, thus
\vspace{-0.3cm}
\be
\Delta M_{[1^n]} = \sum_{m=1}^{n-1}     M_{[1^m]}\otimes M_{[1^{n-m}]}
\ee
For more complicated Young diagrams we have:
\be
\!\!\!\!\!\!\!\!
\Delta M_{[2,1^{n-2}]} =
& \!\!\!\!
\sum_{i=1}^{n-2} \Big(M_{[2,1^{n-2-i}]}\otimes M_{[1^i]}
+ M_{[1^i]}\otimes M_{[2,1^{n-2-i}]}\Big)
+ \sum_{\stackrel{i,j\geq 1}{i+j=n}}
\frac{(1-q)(1-q^2t^{n-2})}{(1-t)(1-qt^{n-1})}
\frac{(1-t^{ i})(1-t^j)}{(1-qt^{i-1})(1-qt^{j-1})}
\cdot M_{[1^i]}\otimes M_{[1^j]}
\nn\\ \nn \\
\!\!\!\!\!\!\!\!
\Delta M_{[3,1^{n-3}]} =
&\sum_{i=1}^{n-3} \Big(M_{[3,1^{n-3-i}]}\otimes M_{[1^i]}
+ M_{[1^i]}\otimes M_{[3,1^{n-3-i}]}\Big)
+ \frac{(1-t)(1-q^2)}{(1-q)(1-qt)}\cdot
\sum_{\stackrel{i,j\geq 0}{i+j=n-4}} M_{[2,1^i]}\otimes M_{[2,1^j]}
+  \nn \\
&+ \frac{ (1-q^2)}{ (1-qt)}\cdot\frac{(1-q^3t^{n-3})}{(1-q^2t^{n-2})}\cdot
\sum_{i=1}^{n-2}
\frac{ (1- t^{i})}{ (1-qt^{i-1})}
\frac{(1-qt^{n-i-1})}{(1-q^2t^{n-i-2}) }
\cdot \Big(M_{[2,1^{n-2-i}]}\otimes M_{[1^i]}+ M_{[1^i]} \otimes M_{[2,1^{n-2-i}]} \Big)
\nn \\
\ldots \nn
\ee

\bigskip

Macdonald polynomials can be obtained by solving these equations,
but, like in the case of Schur functions,
the $p$-linear terms are zero-modes of $\Delta$ and
the coefficients in front of them are not fixed by the recursion.
To cure this problem one can impose orthogonality restriction,
but the rule (\ref{duaSchur}) should also be deformed:
\be
\hat M_R := M_{R}\left\{k\cdot\frac{1-q^k}{1-t^k}\frac{\p}{\p p_k}\right\}
\label{duaMcD}
\ee
Alternatively one can rescale time-variables and polynomials so that
\be
\boxed{
{\cal M}_{[n]}= S_{[n]} \left\{\frac{(1-q)^k}{1-q^k}\cdot p_k\right\}, \ \ \ \ \ \ \
{\cal M}_{[1^n]} =  S_{[1^n]} \left\{\frac{(1-t^{-1})^k}{1-t^{-k}}\cdot p_k\right\}
}
\label{macsimpl}
\ee
get expressed through the Schur functions.
Other polynomials, however, remain somewhat more involved --
what is natural, because they should somehow carry the information about the
third combination like $\frac{(1-t/q)^k}{1-(t/q)^{k}}\cdot p_k$,
somehow mixed with the projection from three dimensions.
We make a few more comments on Macdonald case in sec.\ref{MAC} below,
but postpone a detailed consideration to a separate publication.
The main goal of the present paper is to describe the very idea
of lifting from the ordinary to plane partitions.

\section{On $3d$ recursion and the $3d$ analogue of Schur functions}

Coming back to sec.\ref{recur},
now we have a formalism, which allows straightforward extension from $2d$ to $3d$,
i.e. from ordinary to plane partitions.
Namely, we can try to study recursion (\ref{recSchur}) with the Young diagrams
$\mu,\nu,\rho$ substituted by plane partitions, and supplement it by a direct
analogue of the rule (\ref{duaSchur}),
\be
\hat{\cal S}_R := {\cal S}_R\left\{k\frac{\p}{\p p_k^{(i)}}\right\}
\label{dua3dSchur}
\ee
In what follows we label the three directions by zero, one and two primes.
Then the single-row or single-column Young diagrams lie in just one of the three
directions, while all other ordinary Young diagrams -- in two.
For ordinary Schur functions parameters $\alpha$, $\beta$ are just
plus/minus unities, and orthogonality conditions are easily satisfied.

Emerging at the level $n$ are the new vectors
$ \vec\alpha_n(\pi)$ in the $n$-dimensional space of
$p_{n}^{(i)}$, $i = 1,\ldots,n$,
which describe the 3-Schur functions for all plane partitions $\pi$
of the size $|\pi|=n$.
They satisfy the most naive recursion rule, which, together with orthogonality,
defines their scalar products through those of $\vec\alpha$
at the previous levels.

\begin{itemize}

\item{Level $2$:}

Recursion implies
\be
\Delta \CHI_{[2]} = \Delta \CHI_{[2]}' = \Delta \CHI_{[2]}'' = \CHI_{[1]}\otimes
\CHI_{[1]}
\ \ \Longrightarrow \ \
\CHI_{[2]} = \frac{\vec \alpha_2 \vec p_2 + p_1^2}{2}, \ \
\CHI_{[2]}' = \frac{\vec \alpha_2' \vec p_2 + p_1^2}{2}, \ \
\CHI_{[2]}'' = \frac{\vec \alpha_2'' \vec p_2 + p_1^2}{2}
\ee
with 2-dimensional vectors $\vec p_2 = (p_2^{(1)},p_2^{(2)})$ and three $\vec\alpha_2$.

Then orthogonality with the standard scalar product
\be
\Big<p_k^{(i)}|p_l^{(j)}\Big> = k\delta_{k,l}\delta_{i,j}
\label{metric}
\ee
and, more generally,
\be
\Big< p_\pi \Big| p_{\pi'} \Big> \ = \
\prod_{k}\prod_{i=1}^k \Big<(p_{k}^{(i)})^{m_{k,i}}\Big| (p_{k}^{(i)})^{m_{k,i}'} \Big>\
= \prod_{k}\prod_{i=1}^k   k^{m_{k,i}}\cdot m_{k,i}! \cdot \delta_{m_{k,i},m_{k,i}'}
\ee
implies for the three vectors $\alpha_2$, $\alpha_2'$, $\alpha_2''$:
\be
\vec\alpha_2\vec\alpha_2' = \vec\alpha_2\vec\alpha_2''=\vec\alpha_2'\vec\alpha_2''= -1
\ee
what means that they form a Mercedes star (i.e. are the roots of affine $\widehat{SL(3)}$)
and the lengths of the vectors are
\be
\vec\alpha_2^2 = {\vec\alpha_2}'^2  ={\vec\alpha_2}''^2 = 2
\ee
Thus
\be
\Big<\CHI_2\Big|\CHI_2\Big> = \frac{\vec\alpha_2^2}{2} +  \frac{1}{2} = \frac{3}{2}
\ee

These 3-Schur functions
\be
{\cal S}_{[2]}^\pm = \frac{p_1^2 -\frac{1}{\sqrt{2}}\tilde p_2
\pm\sqrt{\frac{3}{2}}\,  p_2}{2}, \ \ \
{\cal S}_{[2]}^0 = \frac{\sqrt{2}\tilde p_2+p_1^2}{2}
\label{3Schurlevel2}
\ee
are the eigenfunctions of the {\bf cut-and-join operator}
\be
\hat{\cal W}_{[2]}^0
= \frac{p_2}{2}\left(-\sqrt{2}\frac{\p}{\p\tilde p_2} + \frac{\p^2}{\p p_1^2}\right)
+ \left(p_1^2-\frac{\tilde p_2}{\sqrt{2}}\right) \frac{\p}{\p p_2}
= p_1^2\frac{\p}{\p p_2} + \frac{ p_2}{2}\frac{\p^2}{\p p_1^2}-
\frac{1}{\sqrt{2}}\Big(\tilde p_2\frac{\p}{\p p_2} + p_2\frac{\p}{\p \tilde p_2}\Big)
\label{W20}
\ee
\be
\hat{\cal W}_{[2]}^0 {\cal S}_{[2]}^0 = 0, \ \ \ \
\hat{\cal W}_{[2]}^0 {\cal S}_{[2]}^\pm = \pm\sqrt{\frac{3}{2}}{\cal S}_{[2]}^\pm
\label{W20ef}
\ee
while the other two operators
\be
\hat{\cal W}_{[2]}^\pm =
p_1^2\Big(-\frac{1}{2}\frac{\p}{\p  p_2}
\mp \frac{\sqrt{3}}{2}\frac{\p}{\p  \tilde p_2}\Big)
+ \frac{-  p_2\pm\sqrt{3}\tilde p_2}{4} \frac{\p^2}{\p p_1^2}
- \frac{1}{\sqrt{2}}\Big(\tilde p_2\frac{\p}{\p p_2} + p_2\frac{\p}{\p \tilde p_2}\Big)
\ee
act as raising and lowering generators:
\be
\hat{\cal W}_{[2]}^+ {\cal S}_{[2]}^+ = 0,  \ \ \ \
\hat{\cal W}_{[2]}^+ {\cal S}_{[2]}^0 = -\sqrt{\frac{3}{2}}{\cal S}_{[2]}^+,
\ \ \ \ \hat{\cal W}_{[2]}^+ {\cal S}_{[2]}^- = \sqrt{\frac{3}{2}}{\cal S}_{[2]}^0
\nn \\
\hat{\cal W}_{[2]}^- {\cal S}_{[2]}^+ = -\sqrt{\frac{3}{2}}{\cal S}_{[2]}^0,  \ \ \ \
\hat{\cal W}_{[2]}^- {\cal S}_{[2]}^0 = \sqrt{\frac{3}{2}}{\cal S}_{[2]}^-,
\ \ \ \ \hat{\cal W}_{[2]}^+ {\cal S}_{[2]}^- = 0
\ee
Two more triples of operators are obtained by $\pm\frac{2\pi}{3}$ rotations
in the $(p_2,\tilde p_2)$ plane.
For example,
\be
\hat{{\cal W}_2^{0}}' = p_1^2\Big(-\frac{1}{2}\frac{\p}{\p p_2} -
\frac{\sqrt{3}}{2}\frac{\p}{\p \tilde p_2}\Big)
- \frac{ p_2+\sqrt{3}\tilde  p_2}{4} \frac{\p^2}{\p p_1^2}
+ \frac{1}{2\sqrt{2}}\left(
(  \sqrt{3}    p_2+\tilde p_2)\frac{\p}{\p   p_2}
+ (p_2-\sqrt{3}  \tilde p_2 ) \frac{\p}{\p\tilde  p_2}
\right)
\label{Woprime}
\ee
also has the three 3-Schur functions as its three eigenfunctions:
\be
\hat{{\cal W}_2^{0}}' {\cal S}_{[2]}^0 = -\sqrt{\frac{3}{2}}{\cal S}_{[2]}^0, \ \ \ \
\hat{{\cal W}_2^{0}}' {\cal S}_{[2]}^+ = 0, \ \ \ \
\hat{{\cal W}_2^{0}}' {\cal S}_{[2]}^- = \sqrt{\frac{3}{2}}{\cal S}_{[2]}^-
\ee
The third  "diagonal" operator
$\hat{{\cal W}_2^{0}}''$
is obtained by changing the signs of all the $\sqrt{3}$ in (\ref{Woprime}).
Covariant description of this Cartanian triple,
in terms of the triple $\vec\alpha_{[2]},\vec\alpha_{[2]}',\vec\alpha_{[2]}''$, is
\be
\hat{\cal W}^{0,\vec\alpha}_{[2]} =
(\vec\alpha'-\vec\alpha'')\left(\Big(p_1^2-\frac{\vec\alpha\vec p_2}{2}\Big)\frac{\p}{\p \vec p_2}
+\frac{\vec p}{2}\Big(\frac{\p^2}{\p p_1^2} - \vec\alpha\frac{\p}{\p \vec p_2}\Big)\right)
\ee
where one substitutes three cyclic permutations of vectors.
These operators commute at the level two, i.e. on the linear space spanned by
$p_2,\tilde p_2$ and $p_1^2$, but at higher levels additional terms
should be taken into account,
see (\ref{W203}) below for the next addition.
The {\bf nine} operators, revealed at the level $2$, are just a tip of an interesting {\bf non-abelian structure},
which looks like non-trivial  generalization of abelian
one in \cite{MMN1}.

\item{Level 3:}

Now we have three
\be
\Delta \CHI_{[3]} = \CHI_{[2]}\otimes \CHI_{[1]} + \CHI_{[1]}\otimes\CHI_{[2]}
\ \ \Longrightarrow \ \
\CHI_{[3]}=\frac{\vec\alpha_3\vec p_3}{3}
+ \frac{(\vec\alpha_2\vec p_2) p_1}{2} + \frac{p_1^3}{6}
\ee
and three
\be
\Delta \CHI_{[21]} = (\CHI_{[2]}'+\CHI_{[2]}'')\otimes \CHI_{[1]}
+ \CHI_{[1]}\otimes(\CHI_{[2]}'+\CHI_{[2]}'')
\ \ \Longrightarrow \ \
\CHI_{[21]} = \frac{\vec\beta_3\vec p_3}{3}
- \frac{(\vec\alpha_2\vec p_2) p_1}{2} + \frac{p_1^3}{3}
\ee
where the convention is that $\CHI_3$ corresponds to a column of length $3$, lying in the
$x$
direction, while $\CHI_{21}$ -- to the Young diagram, lying in orthogonal plane
$(x',x'')$.
The two triples of $3d$ vectors,
$\vec\alpha_3, \vec\alpha_3', \vec\alpha_3''$ and $\vec\beta_3, \vec\beta_3',
\vec\beta_3''$
are now two triples of $3d$ vectors,
satisfying orthogonality conditions
\be
\CHI_3\bot \CHI_3': &
\frac{\vec\alpha_3\vec\alpha'_3}{3} + \frac{\vec\alpha_2\vec\alpha'_2}{2} + \frac{6}{6^2} =
0
\ \ &\Longrightarrow \ \  \vec\alpha_3\vec\alpha'_3 = 1 \nn\\
\CHI_{21}\bot \CHI_{21}': &
\frac{\vec\beta_3\vec\beta'_3}{3} + \frac{\vec\alpha_2\vec\alpha'_2}{2} + \frac{6}{3^2} =
0
\ \ &\Longrightarrow \ \  \vec\beta_3\vec\beta'_3 = -\frac{1}{2} \nn\\
\CHI_3\bot \CHI_{21}: &
\frac{\vec\alpha_3\vec\beta_3}{3} - \frac{\vec\alpha_2^2 }{2} + \frac{6}{6\cdot 3} = 0
\ \ &\Longrightarrow \ \  \vec\alpha_3\vec\beta_3 = 2 \nn\\
\CHI_3\bot \CHI_{21}': &
\frac{\vec\alpha_3\vec\beta'_3}{3} - \frac{\vec\alpha_2\vec\alpha'_2}{2} + \frac{6}{6\cdot
3} = 0
\ \ &\Longrightarrow \ \  \vec\alpha_3\vec\beta'_3 = -\frac{5}{2}
\ee
If we now parameterize the six vectors by
\be
\vec\alpha_3: & (u,2x,0 ),\  (u,-x,x\sqrt{3} ), \ (u,-x,-x\sqrt{3} ), \ \ \
\vec\beta_3:  & (v,2y,0 ),\  (v,-y,y\sqrt{3} ), \ (v,-y,-y\sqrt{3} )
\label{albe3}
\ee
then we get:
\be
-2x^2+u^2 = 1, \ \ \ -2y^2+v^2=-\frac{1}{2}, \ \ \ 4xy+uv = 2, \ \ \
-2xy+uv = -\frac{5}{2} \nn \\
\Longrightarrow \ \ \ xy = \frac{3}{4}, \ \ uv = -1,
-2x^2v^2-2y^2u^2+\frac{9}{4}+1=-\frac{1}{2}
\nn \\
16 \,\frac{x^2}{u^2}+9\,\frac{u^2}{x^2}  =30 \ \
\Longrightarrow \ \  u^2=\frac{8}{3}x^2
\nn \\
\Longrightarrow \ \ \ \ \
x^2 = \frac{3}{2}, \ \ \ y^2=\frac{3}{8}, \ \ \ u^2=4, \ \ \ v^2= \frac{1}{4}
\ee
and the vector lengths are:
\be
\vec\alpha_3^2 = 4x^2+u^2 = 6 + 4 =10, \ \ \ \ \ \ \ \ \ \ \ \
\vec\beta_3^2 = 4y^2+v^2 = \frac{3}{2}+\frac{1}{4} = \frac{7}{4}
\ee
so that
\be
{\cal S}_{[3]}^0 = \frac{2P_3+ \sqrt{6}\tilde p_3}{3}+\frac{\tilde p_2p_1}{\sqrt{2}}
+\frac{p_1^3}{6}, \ \ \ \
{\cal S}_{[3]}^\pm = \frac{4P_3-\sqrt{6}\tilde p_3\pm 3\sqrt{2}  p_3}{6}
+\frac{(- \tilde p_2\pm \sqrt{3 }  p_2)p_1}{2\sqrt{2}}
+\frac{p_1^3}{6},
\nn \\
{\cal S}_{[21]}^0=  \frac{-P_3+ \sqrt{6}\tilde p_3}{6}-\frac{\tilde p_2p_1}{\sqrt{2}}
 +\frac{p_1^3}{3}, \ \ \ \
{\cal S}_{[21]}^\pm= \frac{-2P_3-\sqrt{6}\tilde p_3\pm 3\sqrt{2} p_3}{12}
-\frac{(- \tilde p_2\pm \sqrt{3 }p_2 )p_1}{2\sqrt{2}}    +\frac{p_1^3}{3}
\label{3Schurlevel3}
\ee
and
\be
\Big<\CHI_3\Big|\CHI_3\Big> = \frac{\alpha_3^2}{3} + \frac{\alpha_2^2}{2}+\frac{1}{6}
= \frac{9}{2}, \ \ \ \ \ \ \ \ \ \ \
\Big<\CHI_{21}\Big|\CHI_{21}\Big> = \frac{\beta_3^2}{3} + \frac{\alpha_2^2}{2}+\frac{2}{3}
= \frac{9}{4}
\ee
This is consistent with the most natural form of Cauchy identity for the 3-Schur functions:
\be
\boxed{
\sum_\pi \frac{{\cal S}_\pi\{p\}{\cal S}_\pi\{p'\}}{<{\cal S}_\pi|{\cal S}_\pi>} =
\exp\left(\sum_n \frac{\vec p_n\vec p_n\!'}{n}\right) =
\exp\left(p_1p_1'+\frac{p_2p_2'+\tilde p_2\tilde p_2'}{2}
+\frac{P_3P_3'+ {p_3p_3'} +\tilde p_3\tilde p_3'}{3} + \ldots\right)
}
\ee

At the new level we get {\bf many  new} {cut-and-join operators},
and  new terms are revealed in the old ones.
More details will be provided elsewhere, here we just mention that
operator $\hat{\cal W}_{[2]}^0$  is now  promoted from (\ref{W20})  to

\be
\hat{\cal W}_{[2]}^0 = p_1^2\frac{\p}{\p   p_2}
+ \frac{  p_2}{2}\frac{\p^2}{\p p_1^2}-
\frac{1}{\sqrt{2}}\Big(\tilde p_2\frac{\p}{\p p_2} + p_2\frac{\p}{\p \tilde p_2}\Big)
+\frac{3(\sqrt{3}   p_3+   p_2 p_1)}{2}\frac{\p}{\p P_3}
-\frac{3(   p_3+\sqrt{3}   p_2 p_1)}{2\sqrt{2}}\frac{\p}{\p \tilde p_3} +
\nn \\
+\frac{3}{2}\left(\sqrt{3} P_3-\frac{\tilde p_3 + \sqrt{3} \tilde p_2 p_1)}{\sqrt{2}}\right)
\frac{\p}{\p p_3}
+\frac{ -\sqrt{3} p_3+ p_2 p_1}{\sqrt{2}}\frac{\p^2}{\p \tilde p_2 \p p_1}
+\left(P_3+\frac{-\sqrt{3}\tilde p_3+ \tilde p_2 p_1}{\sqrt{2}}\right)
\frac{\p^2}{\p  p_2\p p_1}
\label{W203}
\ee
and, in addition to (\ref{W20ef}), it now has six new eigenfunctions
(\ref{3Schurlevel3}):
\be
\hat{\cal W}_{[2]}^0 {\cal S}_{[3]}^0 = 0, \ \ \ \
\hat{\cal W}_{[2]}^0 {\cal S}_{[3]}^\pm = \pm 3\sqrt{\frac{3}{2}}{\cal S}_{[3]}^\pm,
\ \ \ \ \ \ \ \ \ \
\hat{\cal W}_{[2]}^0 {\cal S}_{[21]}^0 = 0, \ \ \ \
\hat{\cal W}_{[2]}^0 {\cal S}_{[21]}^\pm = \mp \frac{3}{2}\sqrt{\frac{3}{2}}{\cal S}_{[21]}^\pm
\ee

\item{Level 4:}
This time we have two triples, made from $[4]$ and $[22]$  in three different directions
(in the case of $[22]$ this is the direction, orthogonal to the plane, to which it
belongs),
plus a six-plet made from $[31]$, which depends on a pair of directions,
plus the first essentially 3d configuration $\Yup$. The total number of relevant vectors
in $4d$ space of $p_4^{(1,2,3,4)}$ is $13$.

From
\be
\Delta\CHI_4 = \CHI_1\otimes\CHI_3 + \CHI_2\otimes\CHI_2+ \CHI_3\otimes \CHI_1 \nn \\
\Delta'\CHI_{31}'' = \CHI_1\otimes(\CHI_3'+\CHI_{21}) + \CHI_2'\otimes\CHI_2'
+ \CHI_2'\otimes\CHI_{2}''+\CHI_{2}''\otimes\CHI_2'+ (\CHI_3'+\CHI_{21})\otimes \CHI_1
\\
\Delta\CHI_{22} = \CHI_1\otimes\CHI_{21} + \CHI_2'\otimes\CHI_2'
+ \CHI_{2}''\otimes\CHI_{2}'' + \CHI_{21}\otimes \CHI_1
\nn
\ee
\be
\!\!\!\!\!\!
\Delta\CHI_{_{\Yup}}
=   \CHI_1\otimes(\CHI_{21}+\CHI'_{21}+\CHI''_{21})
+\CHI_2\otimes(\CHI_2'+\CHI_2'') + \CHI_2'\otimes(\CHI_2+\CHI_2'')
+\CHI_2''\otimes(\CHI_2+\CHI_2') + (\CHI_{21}+\CHI'_{21}+\CHI''_{21})\otimes \CHI_1
\nn\ee
where $\CHI_4$ corresponds to the 1-column Young diagram lying along the $x$ axis,
i.e. belonging to either of the two planes, $(x,x')$ or $(x,x'')$,
while ${'\CHI_{31}''}$ and $\CHI_{22}$ are the ordinary Young diagrams lying in the
plane $(x',x'')$, with the leg of length $3$ in the former case along $x'$ and
that of length $2$ along $x''$, so that ${'\CHI_{211}''} = {''\CHI_{31}'}$,
we get:
\be
\CHI_4 = \frac{(\vec\alpha_4\vec p_4)}{4} + \frac{(\vec\alpha_3 \vec p_3)p_1}{3}
+ \frac{(\vec\alpha_2 \vec p_2)^2}{8} + \frac{(\vec\alpha_2\vec p_2)p_1^2}{4}
+ \frac{p_1^4}{24} \nn \\
{'\CHI_{31}''} = \frac{('\!\vec\beta_4''\vec p_4)}{4}
+ \frac{\big((\vec\alpha_3' +\vec\beta_3)\vec p_3\big)p_1}{3}
+ \frac{(\vec\alpha_2' \vec p_2)
\big((\vec\alpha_2'  +2\vec\alpha_2'')\vec p_2)\big)}{8}
+ \frac{\big((2\vec\alpha_2'+\vec\alpha_2'')\vec p_2\big)p_1^2}{4}
+ \frac{p_1^4}{8} \nn \\
\CHI_{22} = \frac{(\vec\gamma_4\vec p_4)}{4} + \frac{(\vec\beta_3 \vec p_3)p_1}{3}
+ \frac{(\vec\alpha'_2 \vec p_2)^2+(\vec\alpha_2''\vec p_2)^2}{8}
- \frac{(\vec\alpha_2\vec p_2)p_1^2}{4}+ \frac{p_1^4}{12} \nn \\
\CHI_{_{\Yup}}
= \frac{(\vec\mu_4 \vec p_4)}{4}
+\frac{\big((\vec\beta_3+\vec\beta_3'+\vec\beta_3'')\vec p_3\big)p_1}{3}
- \frac{(\vec\alpha_2\vec p_2)^2 +
(\vec\alpha_2'\vec p_2)^2+ (\vec\alpha_2''\vec p_2)^2}{8} + \frac{p_1^4}{4}
\ee
Orthogonality conditions now imply:
\be
\CHI_4\bot \CHI_4': & \frac{\vec\alpha_4\vec\alpha'_4}{4}
+ \frac{\vec\alpha_3\vec\alpha'_3}{3}
+\frac{(\vec\alpha_2\vec\alpha'_2)^2}{8}
+\frac{\vec\alpha_2\vec\alpha'_2}{4}
+\frac{1}{24} = 0
& \vec\alpha_4\vec\alpha'_4 = -1
\nn \\
\CHI_4\bot \CHI_{31}': & \frac{\vec\alpha_4\vec\beta'_4}{4}
+ \frac{\vec\alpha_3(\vec\alpha_3+\beta_3'')}{3}
+\frac{(\vec\alpha_2^2)\big(\vec\alpha_2^2+2(\vec\alpha_2\alpha'_2)\big)}{8}
+\frac{2\vec\alpha_2^2+(\vec\alpha_2\vec\alpha'_2)}{4}
+\frac{1}{8} = 0
& \vec\alpha_4\vec\beta'_4 = -\frac{27}{2}
\nn \\
\CHI_4\bot{' \CHI_{31}}: & \frac{\vec\alpha_4{\, '\!\vec\beta}_4}{4}
+ \frac{\vec\alpha_3(\vec\alpha'_3+\beta_3'')}{3}
+\frac{(\vec\alpha_2\vec\alpha_2')
\big((\vec\alpha_2\vec\alpha_2')+2\vec\alpha_2^2 \big)}{8}
+\frac{2(\vec\alpha_2\vec\alpha'_2)+\vec\alpha_2^2}{4}
+\frac{1}{8} = 0
& \vec\alpha_4{\, '\!\vec\beta}_4 = 3
\nn \\
\CHI_4\bot {'\CHI_{31}''}: & \frac{\vec\alpha_4{\,'\!\vec\beta_4''}}{4}
+ \frac{\vec\alpha_3(\vec\alpha'_3+\vec\beta_3)}{3}
+\frac{(\vec\alpha_2\vec\alpha'_2)\big((\vec\alpha_2\vec\alpha'_2)
+2(\vec\alpha_2\vec\alpha''_2)\big)}{8}
+\frac{2(\vec\alpha_2\vec\alpha'_2)+(\vec\alpha_2\vec\alpha''_2)}{4}
+\frac{1}{8} = 0
&  \vec\alpha_4{\,'\!\vec\beta_4''} = -3
\nn \\
\CHI_4\bot \CHI_{22}: & \frac{\vec\alpha_4\vec\gamma_4}{4}
+ \frac{\vec\alpha_3\vec\beta_3}{3}
+\frac{(\vec\alpha_2\vec\alpha'_2)^2+(\vec\alpha_2\vec\alpha''_2)^2}{8}
-\frac{\vec\alpha_2^2}{4}+\frac{1}{12} = 0
&  \vec\alpha_4\vec\gamma_4= -2
\nn \\
\CHI_4\bot \CHI_{22}': & \frac{\vec\alpha_4\vec\gamma'_4}{4}
+ \frac{\vec\alpha_3\vec\beta'_3}{3}
+\frac{(\vec\alpha_2^2)^2+(\vec\alpha_2\vec\alpha''_2)^2}{8}
-\frac{\vec\alpha_2\vec\alpha_2'}{4} +\frac{1}{12} = 0
&  \vec\alpha_4\vec\gamma_4' = -\frac{1}{2}
\nn \\
\CHI_4\bot \CHI_{_{\Yup}}: & \frac{\vec\alpha_4\vec\mu_4}{4}
+ \frac{\vec\alpha_3(\vec\beta_3+\vec\beta_3'+\vec\beta_3'')}{3}
-\frac{(\vec\alpha_2^2)^2+(\vec\alpha_2\vec\alpha'_2)^2+(\vec\alpha_2\vec\alpha''_2)^2}{4}
+\frac{1}{4} = 0
&  \vec\alpha_4\vec\mu_4 = 6
\nn
\ee
\be
'\!\CHI_{31}\bot \CHI_{31}': &
\frac{'\!\vec\beta_4\vec\beta'_4}{4}
+ \frac{(\vec\alpha_3'+\vec\beta_3'')(\vec\alpha_3+\vec\beta_3'' )}{3}
+\frac{(\vec\alpha_2\vec\alpha_2')
\big((\vec\alpha_2+2\vec\alpha'_2)(\vec\alpha_2'+2\vec\alpha_2)\big)
+\big(\vec\alpha_2(\vec\alpha_2'+2\vec\alpha_2)\big)
\big(\vec\alpha_2'(\vec\alpha_2+2\vec\alpha'_2)\big)}{16}+
& '\!\vec\beta_4\vec\beta'_4 = -3
\nn\\
&+\frac{(2\vec\alpha_2+\vec\alpha'_2)(2\vec\alpha_2'+\vec\alpha_2)}{4}
+\frac{3}{8} = 0
\nn\\
\CHI_{31}'\bot \CHI_{31}'': &
\frac{\vec\beta_4'\vec\beta''_4}{4}
+ \frac{(\vec\alpha_3+\vec\beta_3'')(\vec\alpha_3+\vec\beta_3' )}{3}
+\frac{\vec\alpha_2^2
\big((\vec\alpha_2+2\vec\alpha'_2)(\vec\alpha_2+2\vec\alpha''_2)\big)
+\big(\vec\alpha_2(\vec\alpha_2+2\vec\alpha'_2)\big)
\big(\vec\alpha_2(\vec\alpha_2+2\vec\alpha''_2)\big)}{16}+
& \vec\beta_4'\vec\beta''_4 =  -\frac{15}{2}
\nn \\
&+\frac{(2\vec\alpha_2+\vec\alpha'_2)(2\vec\alpha_2+\vec\alpha''_2)}{4}
+\frac{3}{8} = 0
\nn\\
'\!\CHI_{31}\bot {\CHI_{31}''}: &
\frac{'\!\vec\beta_4 \vec\beta''_4}{4}
+ \frac{(\vec\alpha'_3+\vec\beta_3'')(\vec\alpha_3+\vec\beta_3' )}{3}
+\frac{(\vec\alpha_2\vec\alpha_2')
\big((\vec\alpha'_2+2\vec\alpha_2)(\vec\alpha_2+2\vec\alpha''_2)\big)
+ \big(\vec\alpha_2(\vec\alpha'_2+2\vec\alpha_2)\big)
\big(\vec\alpha_2'(\vec\alpha_2+2\vec\alpha''_2)\big)}{16}+\ \ \
& '\!\vec\beta_4\vec\beta''_4 = 3
\nn \\
& +\frac{(2\vec\alpha_2'+\vec\alpha_2)(2\vec\alpha_2+\vec\alpha''_2)}{4}
+\frac{3}{8} = 0
\nn\\
'\!\CHI_{31}\bot {''\CHI_{31}}: &
\frac{'\!\vec\beta_4{''\!\vec\beta}_4}{4}
+ \frac{(\vec\alpha_3'+\vec\beta_3'')(\vec\alpha''_3+\vec\beta_3' )}{3}
+\frac{(\vec\alpha'_2\vec\alpha''_2)
\big((\vec\alpha'_2 +2\vec\alpha_2 )(\vec\alpha''_2+2\vec\alpha_2 )  \big)
+\big(\vec\alpha_2''(\vec\alpha'_2 +2\vec\alpha_2 )\big)
\big(\vec\alpha_2'(\vec\alpha''_2+2\vec\alpha_2 )  \big)  }{16} +\ \ \ \
& '\!\vec\beta_4{''\!\vec\beta}_4 =  -3
\nn\\
& +\frac{(2\vec\alpha'_2 + \vec\alpha_2 )(2\vec\alpha''_2+ \vec\alpha_2 )}{4}
+\frac{3}{8} = 0
\nn
\ee
\be
\CHI_{31}'\bot \CHI_{22}: & \frac{\vec\beta_4'\vec\gamma_4}{4}
+ \frac{(\vec\alpha_3+\vec\beta_3'')\vec\beta_3}{3}
+\frac{(\vec\alpha_2\vec\alpha_2')\big((\vec\alpha_2+2\vec\alpha'_2)\vec\alpha_2'\big)
+(\vec\alpha_2\vec\alpha_2'')\big((\vec\alpha_2+2\vec\alpha'_2)\vec\alpha_2''\big)}{8}
-\frac{(2\vec\alpha_2+\vec\alpha_2')\vec\alpha_2}{4}+\frac{1}{4} = 0
&  \vec\beta_4'\vec\gamma_4 = 0
\nn \\
'\!\CHI_{31}\bot \CHI_{22}: & \frac{'\!\vec\beta_4\vec\gamma_4}{4}
+ \frac{(\vec\alpha_3'+\vec\beta_3'')\vec\beta_3}{3}
+\frac{(\vec\alpha_2'\vec\alpha_2')\big((\vec\alpha_2'+2\vec\alpha_2)\vec\alpha_2'\big)
+(\vec\alpha_2'\vec\alpha_2'')\big((\vec\alpha_2'+2\vec\alpha_2)\vec\alpha_2''\big)}{8}
-\frac{(2\vec\alpha'_2+\vec\alpha_2)\vec\alpha_2}{4}+\frac{1}{4} = 0
&  '\!\vec\beta_4\vec\gamma_4 = \frac{3}{2}
\nn \\
'\!\CHI_{31}''\bot \CHI_{22}: & \frac{'\!\vec\beta_4''\vec\gamma_4}{4}
+ \frac{(\vec\alpha_3'+\vec\beta_3)\vec\beta_3}{3}
+\frac{(\vec\alpha_2'\vec\alpha_2')\big((\vec\alpha_2'+2\vec\alpha''_2)\vec\alpha_2'\big)
+(\vec\alpha_2'\vec\alpha_2'')\big((\vec\alpha_2'+2\vec\alpha''_2)\vec\alpha_2''\big)}{8}
-\frac{(2\vec\alpha'_2+\vec\alpha_2'')\vec\alpha_2}{4}+\frac{1}{4} = 0
& '\!\vec\beta_4''\vec\gamma_4 = -\frac{3}{2}
\nn \\
\CHI_{31}'\bot \CHI_{_{\Yup}}: & \frac{\vec\beta_4'\vec\mu_4}{4}
+ \frac{(\vec\alpha_3+\vec\beta_3'')(\vec\beta_3+\vec\beta_3'+\vec\beta_3'')}{3}
-\frac{(\vec\alpha_2\vec\alpha_2)\big((\vec\alpha_2+2\vec\alpha'_2)\vec\alpha_2\big)
+(\vec\alpha_2\vec\alpha_2')\big((\vec\alpha_2+2\vec\alpha'_2)\vec\alpha_2'\big)
+(\vec\alpha_2\vec\alpha_2'')\big((\vec\alpha_2+2\vec\alpha'_2)\vec\alpha_2''\big)
}{4}
+\frac{3}{4} = 0
&  \vec\beta_4'\vec\mu_4 = 0
\nn \\
\nn \\
\CHI_{22}\bot \CHI_{22}': &  \frac{\vec\gamma_4\vec\gamma'_4}{4}
+ \frac{\vec\beta_3\vec\beta'_3}{3}
+\frac{(\vec\alpha_2\vec\alpha'_2)^2+(\vec\alpha_2\vec\alpha''_2)^2
+ (\vec\alpha'_2\vec\alpha''_2)^2  +({{ \alpha_2''}}^2)^2}{8}
+\frac{\alpha_2\vec\alpha_2'}{4}+\frac{1}{6} = 0
&  \vec\gamma_4\vec\gamma'_4 = -\frac{5}{2}
\nn \\
\nn \\
\CHI_{22}\bot \CHI_{_{\Yup}}: & \frac{\vec\gamma_4\vec\mu_4}{4}
+ \frac{\vec\beta_3(\vec\beta_3+\vec\beta_3'+\vec\beta_3'')}{3}
-\frac{(\vec\alpha_2\vec\alpha'_2)^2+ (\vec\alpha_2\vec\alpha''_2)^2
+2(\vec\alpha'_2\vec\alpha''_2)^2
+({\alpha'_2}^2)^2+  ({\alpha''_2})^2}{8}
+\frac{1}{2} = 0
& \vec\gamma_4\vec\mu_4= 3
\nn
\ee

We can try the following ansatz for 13 vectors in $4d$ space:
\be
\vec\mu_4 = ( s^{-1},0,0,0)
\nn \\
\vec\alpha_4 = (6s,u,2x,0),\
\vec\alpha_4' = (6s,u,-x,x\sqrt{3}),\
\vec\alpha_4'' = (6s,u,-x,-x\sqrt{3})
\nn \\
{'\!\vec\beta_4''} = (0, \epsilon, 2y,0), \ \ ''\!\vec\beta_4' = ( 0, -\epsilon,-2y, 0), \nn\\
{''\!\vec\beta_4} = ( 0, \epsilon, -y\sqrt{3}, -y), 
\ \  \vec\beta_4'' = (0,-\epsilon, y\sqrt{3}, y),
\nn\\
 \vec\beta_4' = ( y\sqrt{3}, -y, v,-3s), \ \ '\!\vec\beta_4 = (-y\sqrt{3}, y, v,-3s),
 \nn\\
\vec\gamma_4 = (3s,w,2z,0 ),\
\vec\gamma_4'=(3s,-w,-z,z\sqrt{3} ),\
\vec\gamma_4''=(3s,w,-z,-z\sqrt{3})
\ee
One can also try other mutual $60^\circ$ rotations in the 2d plane in the last two
coordinates.
Unfortunately, this time the miracle does not repeat: there is no immediate
solution, at least in this form.

\item{Level 5:} We will have six triples, made from $[5],[41],[32],[311],[221],[2111]$
plus two triples of  $3d$
configurations $<2,1,1>$ ($[2]$ atop $[21]$) and $1$ atop $[2,2]$.
The total number of vectors will in $5d$ space is $24$.

\item{Level 6:} $48$ vectors in $6d$ space.

\item{Level 7:} $86$ vectors in $7d$ space -- and so on, in accordance with (\ref{MMah}).

\end{itemize}

\newpage

\section{Back from plane to ordinary Schur functions}

One can consider elimination of extra $p_k^{(i)}$-variables by a kind of
projection/contraction
onto a 1-dimensional line/direction  in $n$-dimensional space.
The choice of this direction should be done separately for each $n$.
Reduction/projection to ordinary Schur functions is nearly trivial:
the relevant line is just the $x$ axis: $p^{(i)}_k=0$ for $i=2,\ldots,n$.


\begin{itemize}

\item{$n=2:$}

In this case we have three polynomials
\be
\S^\bot_2 = \S_2^{0}=\frac{\sqrt{2} \tilde p_2  + p_1^2}{2}, \ \ \ \ \ \ \
\S_2^{\pm} = \frac{ \pm \sqrt{\frac{3}{2}}p_2  - \frac{1}{\sqrt{2}}\tilde p_2 + p_1^2}{2 }
\ee
and we want to eliminate $\tilde p_2=p_2^{(2)}$.
Just putting $p_2^{(2)}=0$ still leaves us with {\it three} polynomials,
moreover even the resulting $S^{\pm}$ differ from the ordinary Schur functions
\be
S_{[2]} = \frac{p_2+p_1^2}{2}=S^+_2, \ \ \ \ \
S_{[11]} = \frac{-p_2+p_1^2}{2} =S^-_2
\label{2Schurlevel2}
\ee
by the coefficient in front of $p_2$.

A viable alternative is to deform the recursion rule:
\be
\Delta \S_2^{\pm} = \S_1\otimes \S_1,   \ \ \ \ \ \ \
\Delta \S_2^\bot = 2h^2 \cdot \S_1\otimes \S_1
\label{hreclevel2}
\ee
what implies
\be
\S_2^\bot = \frac{\vec\alpha_2^\bot \vec p_2 + 2h^2p_1^2}{2}, \ \ \ \ \ \ \ \ \
\S_2^{\pm} = \frac{\vec\alpha_2^\pm \vec p_2 + p_1^2}{2}
\ee
If we do not deform orthogonality condition, then
$\Big<p_2^{(a)}\Big|p_2^{(b)}\Big>\ = 2\delta^{ab}$,
\be
\vec\alpha_2^+\vec\alpha_2^-=-1, \ \ \ \ \ \vec\alpha_2^\bot\vec\alpha_2^{\pm}=-2h^2
\ee
If we now also {\it preserve} the relation
\be
\vec\alpha_2^++\vec\alpha_2^-+\vec\alpha_2^\bot = 0
\label{zerosum}
\ee
then
\be
\vec\alpha_2^\bot = (2h ,0), \ \ \ \vec\alpha^{\pm}_2 = (-h,\pm\sqrt{1+h^2})
\ \ \ \ \ \ \Longrightarrow \ \ \ \ (\vec\alpha_2^\bot)^2=4h^2, \ \ \
(\vec\alpha^{\pm}_2)^2 = 1+2h^2
\ee
and we obtain a smooth interpolation between the cases
(\ref{3Schurlevel2}) of plain-partition at $h=1/\sqrt{2}$
and (\ref{2Schurlevel2}) ordinary Schur functions at $h=0$:
\be
\S_2^\bot = \frac{2h \tilde p_2 + 2h^2p_1^2}{2}, \ \ \ \ \ \ \ \ \
\S_2^{\pm} = \frac{-h \tilde p_2\pm \sqrt{1+h^2}\,  p_2 + p_1^2}{2}
\label{h2Schurlevel2}
\ee
In the latter case $\S_2^\bot$ is naturally vanishing and "disappears".

The $h$-deformation of cut-and-join operator
\be
\hat{\cal W}_{[2]}^0(h) = p_1^2\frac{\p}{\p   p_2}
+ \frac{  p_2}{2}\frac{\p^2}{\p p_1^2}-
h\Big(\tilde p_2\frac{\p}{\p p_2} + p_2\frac{\p}{\p \tilde p_2}\Big)
+ \ldots
\label{W20h}
\ee
smoothly interpolates between (\ref{W20}) and (\ref{W2Schur}).
Its eigenfunctions at level 2 are
\be
{\cal S}^\lambda_{[2]}(h) \sim
{(1-\lambda^2)\tilde p_2+ h\lambda p_2+hp_1^2}
\label{Sh}
\ee
with eigenvalues $\lambda$, which are the three roots of the characteristic
equation
\be
\lambda(\lambda^2-h^2-1)=0
\ee
Normalization of eigenfunctions is dictated by the recursion rule,
and imposing (\ref{hreclevel2}) we obtain (\ref{h2Schurlevel2}).

\newpage

\item{$n=3:$}

The crucial fact here is that $\vec\alpha_3 \sim \vec\beta_3'+\vec\beta_3''$
so that they lie in a plane and can vanish together after projection to orthogonal line.
Indeed, according to (\ref{albe3}),
$\vec\alpha_3= (u,2x,0 )=(2,\sqrt{6},0)$
and
$\beta_3'+\beta_3''=(2v,-2y,0)=(-1,-\sqrt{3/2},0)=-\frac{1}{2}\vec\alpha_3$.

Deformed recursion
\be
\Delta \S_{[3]}^\bot = 2h^2\Big(\S_{[2]}^\bot\otimes \S_{[1]} +
\S_{[1]}\otimes\S_{[2]}^\bot\Big)
\ \ \Longrightarrow \ \
\S_{[3]}^\bot=\frac{\vec\alpha_3^\bot\vec p_3}{3}
+ 2h^2\frac{(\vec\alpha_2^\bot\vec p_2) p_1}{2} + 4h^4\frac{p_1^3}{6}
\nn\\
\Delta \S_{[3]}^\pm =  \S_{[2]}^\pm\otimes \S_{[1]} + \S_{[1]}\otimes\S_{[2]}^\pm
\ \ \Longrightarrow \ \
\S_{[3]}^\pm=\frac{\vec\alpha_3^\pm\vec p_3}{3}
+ \frac{(\vec\alpha_2^\pm\vec p_2) p_1}{2} + \frac{p_1^3}{6}
\ee
and
\be
\Delta \S_{[21]}^\bot = (\S_{[2]}^++\S_{[2]}^-)\otimes \S_{[1]}
+ \S_{[1]}\otimes(\S_{[2]}^++\S_{[2]}^-)
\ \ \Longrightarrow \ \
\S_{[21]}^\bot = \frac{\vec\beta_3^\bot\vec p_3}{3}
+ \frac{\big((\vec\alpha_2^++\vec\alpha_2^-)\vec p_2\big) p_1}{2} + \frac{p_1^3}{3}
\nn \\
\Delta \S_{[21]}^\pm = \Big(\S_{[2]}^\bot+2h^2\S_{[2]}^\mp\Big)\otimes \S_{[1]}
+ \S_{[1]}\otimes\Big(\S_{[2]}^\bot+2h^2\S_{[2]}^\mp\Big)
\ \ \Longrightarrow \ \ \ \ \ \ \ \ \ \ \ \ \ \ \ \ \ \ \ \ \nn \\
\Longrightarrow \ \
\S_{[21]}^\pm = \frac{\vec\beta_3^\pm\vec p_3}{3}
+ \frac{\big((\vec\alpha_2^\bot+2h^2\vec\alpha_2^\mp)\vec p_2\big) p_1}{2} +
2h^2\frac{p_1^3}{3}
\ee
leads to deformed orthogonality conditions:
\be
\CHI_3^\bot \bot \CHI_3^\pm: &
\frac{\vec\alpha_3^\bot\vec\alpha_3^\pm}{3} +
2h^2\frac{\vec\alpha_2^\bot\vec\alpha_2^\pm}{2}
+ 4h^4\frac{6}{6^2} = 0
\ \ &\Longrightarrow \ \  \vec\alpha_3^\bot\vec\alpha_3^\pm = 4h^4
\ \stackrel{h^2=\frac{1}{2}}{\longrightarrow} \ 1
\nn\\
\CHI_3^+ \bot \CHI_3^-: &
\frac{\vec\alpha_3^+\vec\alpha_3^-}{3} + \frac{\vec\alpha_2^+\vec\alpha_2^-}{2}
+ \frac{6}{6^2} = 0
\ \ &\Longrightarrow \ \  \vec\alpha_3^+\vec\alpha_3^- = \boxed{1}
\ \stackrel{h^2=\frac{1}{2}}{\longrightarrow} \ 1
\nn\\
\CHI_{21}^\bot \bot \CHI_{21}^\pm: &
\frac{\vec\beta_3^\bot\vec\beta_3^\pm}{3}
+ \frac{(\vec\alpha_2^++\vec\alpha_2^-)(\vec\alpha_2^\bot+2h^2\vec\alpha_2^\pm)}{2}
+ 2h^2\frac{6}{3^2} = 0
\ \ &\Longrightarrow \ \  \vec\beta_3^\bot\vec\beta_3^\pm = 2h^2-6h^4
\ \stackrel{h^2=\frac{1}{2}}{\longrightarrow} \ -\frac{1}{2}
\nn\\
\CHI_{21}^+ \bot \CHI_{21}^-: &
\frac{\vec\beta_3^+\vec\beta_3^-}{3}
+ \frac{(\vec\alpha_2^\bot+2h^2\vec\alpha_2^-)(\vec\alpha_2^\bot+2h^2\vec\alpha_2^+)}{2}
+ 4h^4\frac{6}{3^2} = 0
\ \ &\Longrightarrow \ \  \vec\beta_3^+\vec\beta_3^- = -6h^2+10h^4
\ \stackrel{h^2=\frac{1}{2}}{\longrightarrow} \ -\frac{1}{2}
\nn\\ \nn \\
\CHI_3^\bot\bot \CHI_{21}^\bot: &
\frac{\vec\alpha_3^\bot\vec\beta_3^\bot}{3} +
2h^2\frac{\vec\alpha_2^\bot(\vec\alpha_2^++\vec\alpha_2^-) }{2}
+ 4h^4\frac{6}{6\cdot 3} = 0
\ \ &\Longrightarrow \ \  \vec\alpha_3^\bot\vec\beta_3^\bot = 8h^4
\ \stackrel{h^2=\frac{1}{2}}{\longrightarrow} \ 2
\nn\\
\CHI_3^\bot\bot \CHI_{21}^\pm: &
\frac{\vec\alpha_3^\bot\vec\beta_3^\pm}{3} +
2h^2\frac{\vec\alpha_2^\bot(\vec\alpha_2^\bot+2h^2\vec\alpha_2^\mp)}{2}
+ 8h^6\frac{6}{6\cdot 3} = 0
\ \ &\Longrightarrow \ \  \vec\alpha_3^\bot\vec\beta_3^\pm = -12h^4+4h^6
\ \stackrel{h^2=\frac{1}{2}}{\longrightarrow} \ -\frac{5}{2}
\nn\\
\CHI_3^\pm\bot \CHI_{21}^\bot: &
\frac{\vec\alpha_3^\pm\vec\beta_3^\bot}{3} +
\frac{\vec\alpha_2^\pm(\vec\alpha_2^++\vec\alpha_2^-) }{2} + \frac{6}{6\cdot 3} = 0
\ \ &\Longrightarrow \ \  \vec\alpha_3^\pm\vec\beta_3^\bot = \boxed{-1}-3h^2
\ \stackrel{h^2=\frac{1}{2}}{\longrightarrow} \ -\frac{5}{2}
\nn\\
\CHI_3^\pm\bot \CHI_{21}^\pm: &
\frac{\vec\alpha_3^\pm\vec\beta_3^\pm}{3} +
\frac{\vec\alpha_2^\pm(\vec\alpha_2^\bot+2h^2\vec\alpha_2^\mp)}{2}
+ 2h^2\frac{6}{6\cdot 3} = 0
\ \ &\Longrightarrow \ \  \vec\alpha_3^\pm\vec\beta_3^\pm = 4h^2
\ \stackrel{h^2=\frac{1}{2}}{\longrightarrow} \ 2
\nn\\
\CHI_3^\pm\bot \CHI_{21}^\mp: &
\frac{\vec\alpha_3^\pm\vec\beta_3^\mp}{3} +
\frac{\vec\alpha_2^\pm(\vec\alpha_2^\bot+2h^2\vec\alpha_2^\pm)}{2}
+ 2h^2\frac{6}{6\cdot 3} = 0
\ \ &\Longrightarrow \ \  \vec\alpha_3^\pm\vec\beta_3^\mp = -2h^2-6h^4
\ \stackrel{h^2=\frac{1}{2}}{\longrightarrow} \ -\frac{5}{2}
\nn\ee
Boxed are the products, surviving at $h=0$,
while arrows show how
symmetric expressions are restored when the deformation parameter $2h^2=1$.

Surviving are also the linear dependencies
\be
(1-h^2)\vec\alpha^\pm_3 + \vec\beta^\bot_3 + \vec\beta^\mp_3 = 0
\ee
and
\be
(3h^2-1)\vec\alpha^\bot_3 + 2h^2(\vec\beta^+_3 + \vec\beta^-_3) = 0
\ee
It follows that
\be
\!\!\!\!
(\vec\alpha_3^\pm)^2 =\frac{(1+2h^2)(1+3h^2)}{1-h^2}, \ \ \ \
(\vec\beta_3^\bot)^2 = 1+3h^4, \ \ \ \
(\vec\alpha_3^\bot)^2 = 16h^6\,\frac{3-h^2}{3h^2-1}, \ \ \ \
(\vec\beta_3^\pm)^2 = 2h^4(5-3h^2)
\ee
Note that these squares are not everywhere positive for real $h^2$.

\end{itemize}

\section{Macdonald polynomials from the 3-Schur functions
\label{MAC}}

To consider {\bf $q,t$-deformations} we should choose other lines to project on.
At level $n=2$
the appropriate line has the slope $\theta$ with
\be
\sin(3\theta) = 2\sqrt{2}\cdot\frac{q-t }{ \sqrt{(1-q^2)(1-t^{2})}}
\label{thetaMac}
\ee
The simplest approach to the study of $q,t$-deformations is through
the algebra of cut-and-join operators.
For example,
\be
\boxed{
\hat{\cal W}^{\theta}_{[2]}(h) =
\frac{ p_2}{2} \p_1^2 + p_1^2  \p_2
- h\cos(3\theta) (\tilde p_2 \p_2+p_2\tilde \p_2)
-\frac{\sin(3\theta)}{\sqrt{2}}(p_2\p_2 - 2h^2\tilde p_2\tilde \p_2 )+\ldots
}
\label{W20thetah}
\ee
interpolates between the 3-Schur operator (\ref{W20}),
\be
\hat {\cal W}_{[2]}^0=\frac{ p_2}{2} \p_1^2 + p_1^2  \p_2
- \frac{1}{\sqrt{2}} (\tilde p_2 \p_2+p_2\tilde \p_2) +\ldots
\ee
at $h=\frac{1}{\sqrt{2}}$ and $\theta=0$
and the {\it differential} Macdonald operator
\be
\widehat{WM}_{[2]}(q,t) =
\frac{ p_2}{2} \p_1^2 + p_1^2  \p_2
-\frac{\sin(3\theta)}{\sqrt{2}}p_2\p_2 + \ldots
\ee
at $h=0$ and $\theta$ from (\ref{thetaMac}).
Like in the case of $3-2$ Schur interpolation (\ref{Sh}),
the eigenfunctions of (\ref{W20thetah}),
\be
{\cal S}_{[2]}^\lambda(h,\theta)\sim
h\cos(3\theta)\Big(
\lambda   p_2 + p_1^2\Big)
+ \left(1-\frac{\sin(3\theta)}{\sqrt{2}}\lambda-\lambda^2\right) \tilde p_2
\ee
with $\lambda$ -- the three roots of characteristic equation
\be
\sqrt{2}\lambda(\lambda^2-1-h^2) + \sin(3\theta)\Big(\lambda^2(1-2h^2)+2h^2\Big) = 0
\ee
interpolate between
the 3-Schur functions (\ref{3Schurlevel2})
and Macdonald polynomials
-- in coordinates which are a further rescaling of $p_2$ in (\ref{macsimpl}):
\be
M_{[2]} = M_2^+=
\frac{1}{2}\left(\sqrt{\frac{(1-q)(1+t )}{(1+q)(1-t )}}\cdot p_2+p_1^2\right),
\ \ \ \ \
M_{[11]}=M_2^- =
\frac{1}{2}\left(-\sqrt{\frac{(1+q)(1-t )}{(1-q)(1+t )}}\cdot p_2 + p_1^2\right)
\ee
In these coordinates the two polynomials are orthogonal in the metric (\ref{metric})
-- as it should be for the eigenfunctions
of the operator (\ref{W20thetah}), which is hermitian in this metric,
since $p_k^\dagger = k\p_k$, $\p_k^\dagger = \frac{1}{k}p_k$ and
$(p_k\p_k)^\dagger = \p_k^\dagger p_k^\dagger = p_k\p_k$.

Actually, at level $2$ only one combination of $q$ and $t$
emerges in the slope (\ref{thetaMac}) -- simply because there
is only one angle in this case,--
but $q$ and $t$ get separated at higher levels $n>2$,
where  projection line   in the $n$-dimensional space
is parameterized by $n-1$ angles.
One can also consider a tripe-$h$ deformation of (\ref{W20thetah})
with three different deformation parameters.

\section{Conclusion}

In this paper we suggested a way to extend the notion of Schur functions
to the case of plane partitions.
The definition which allows such generalization is in terms of recursion
in the size of the Young diagrams.
Postulating it in the most naive form we get a $3d$-symmetric analogue of
Schur functions, while deformation of the coefficients in the recursion
allows to go back from $3d$ to $2d$.
Moreover, the freedom in this deformation seems sufficient to provide
reductions not only to the ordinary Schur functions, but also to Macdonald
polynomials.
After such functions are constructed and investigated,
one can study their network-model averages and, hopefully,
get the corresponding extension of the superintegrability relation
(\ref{charchar}).
Practical realization of this program is, however, quite tedious and
will be further developed elsewhere.
The present paper just describes  the main idea and shows some miracles,
confirming that the idea can work.

\section*{Acknowledgements}

I am indebted for discussion of the 3d generalizations of characters
to H.Awata, H.Kanno, A.Mironov and especially to Y.Zenkevich.

This work was performed at the Institute for Information Transmission Problems
with the financial support
of the Russian Science Foundation (Grant No.14-50-00150).

\end{document}